\documentclass[sigconf]{acmart}
\AtBeginDocument{%
  \providecommand\BibTeX{{%
    \normalfont B\kern-0.5em{\scshape i\kern-0.25em b}\kern-0.8em\TeX}}}

\setcopyright{acmcopyright}
\copyrightyear{2020}
\acmYear{2020}
\setcopyright{rightsretained}
\acmConference[NICE '20]{Neuro-inspired Computational Elements Workshop}{March 
17--20, 2020}{Heidelberg, Germany}
\acmBooktitle{Neuro-inspired Computational Elements Workshop (NICE '20), March 
17--20, 2020, Heidelberg, Germany}\acmDOI{10.1145/3381755.3381767}
\acmISBN{978-1-4503-7718-8/20/03}

\newcommand{\cm}{c_\text{m}}

\newcommand{\EL}{E_\text{L}}

\newcommand{\gLsoma}{g_\text{L}^\text{s}}

\newcommand{\gtotden}{g^\text{d}}
\newcommand{\gtotsoma}{g^\text{s}}
\newcommand{\gtotsomai}[1]{g^{\text{s},i}}
\newcommand{\rate}{{\bf r}}

\newcommand{\usoma}{u^\text{s}}
\newcommand{\ut}{u^\text{t}}
\newcommand{\uti}[1]{u^{\text{t},i}}

\newcommand{\utotden}{\bar u^\text{d}}
\newcommand{\utotsoma}{\bar u^\text{s}}
\newcommand{\utotsomai}[1]{\bar u^{\text{s},#1}}
\newcommand{\w}{{\bf W}} % custom macros

\begin{document}

\title{Conductance-based dendrites perform reliability-weighted opinion pooling}

\author{Jakob Jordan, Mihai A. Petrovici, Walter Senn}
\email{jordan,petrovici,senn@pyl.unibe.ch}
% \author{Mihai A. Petrovici}
% \email{petrovici@pyl.unibe.ch}
% \author{Walter Senn}
% \email{senn@pyl.unibe.ch}
\affiliation{%
  % \institution{Department of Physiology}
  \institution{University of Bern}
  % \streetaddress{P.O. Box 1212}
  \city{Bern}
  \state{Switzerland}
}

\author{Jo\~ao Sacramento}
\email{sacramento@ini.ethz.ch}
\affiliation{
  \institution{UZH / ETH}
  \city{Z\"urich}
  \state{Switzerland}
}



\begin{abstract}
Cue integration, the combination of different sources of information to reduce uncertainty, is a fundamental computational principle of brain function.
Starting from a normative model we show that the dynamics of multi-compartment neurons with conductance-based dendrites naturally implement the required probabilistic computations.
The associated error-driven plasticity rule allows neurons to learn the relative reliability of different pathways from data samples, approximating Bayes-optimal observers in multisensory integration tasks.
Additionally, the model provides a functional interpretation of neural recordings from multisensory integration experiments and makes specific predictions for membrane potential and conductance dynamics of individual neurons.
\end{abstract}





\maketitle

\section{Introduction}
Animals need to operate successfully in their environment based on sensory information and prior expectations that are both incomplete and uncertain.
To overcome the limitations of individual information sources it is useful to combine them, for example sensory inputs with prior expectations, sensory inputs from different modalities or the information from different receptive fields.
A probabilistic model of cue integration shows that combining multiple sources of information indeed reduces uncertainty.
However, to do so successfully requires knowledge about the reliability of each source: the maximum-a-posteriori (MAP) estimate is a linear combination of individual cues, each weighted with their relative reliability \citep{knill2004bayesian}.
Behavioral evidence \citep{ernst2002humans,fetsch2009dynamic,nikbakht2018supralinear} demonstrates that humans and non-human animals indeed are able to optimally integrate multisensory stimuli to improve their performance compared to unisensory testing conditions.
What kind of neural circuitry enables these probabilistic computations?
We propose that multi-compartment neuron models with conductance-based dendrites are naturally equipped to learn the reliability of different sensory streams and to use this information to perform approximately optimal cue integration.
Neuron and synapse dynamics are jointly derived from an energy-minimization principle.
The resulting neuron dynamics coincide with standard leaky integrators with multiple dendritic compartments. The associated plasticity rule is reminiscent of error-driven learning rules \citep{widrow1960adaptive,urbanczik2014learning}, but contains an additional term to learn the relative reliabilities of different pathways.
To illustrate the model, we train it on a multisensory integration task and demonstrate that it can approximate Bayes-optimal inference.
Furthermore, the dynamics of the trained model is in good agreement with experimental findings and allows us to make specific predictions on membrane potentials and conductances in multisensory integration experiments.
Our model connects a normative approach to cue integration with circuit-level implementations, bridging the scales from behaviour to individual neurons.

\begin{figure*}
  \centering
  \includegraphics[width=1.\linewidth]{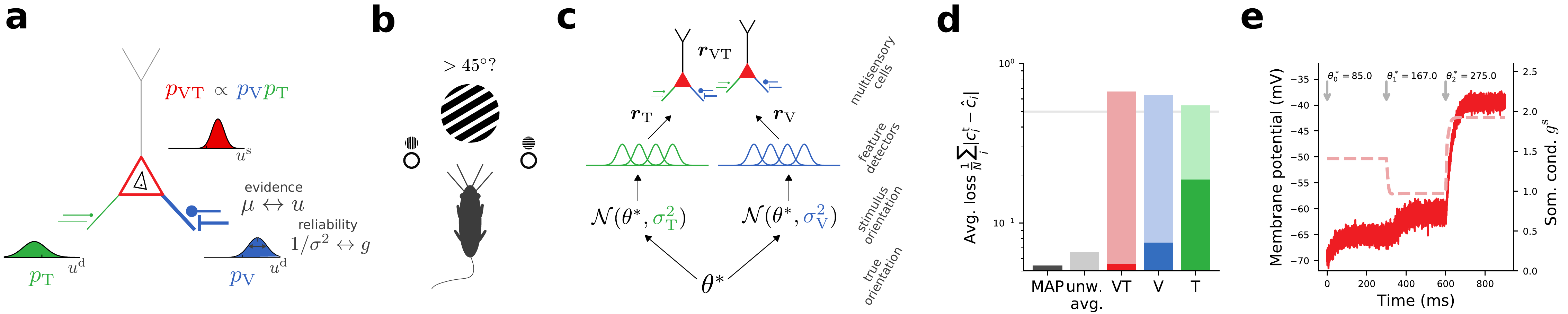}
  \caption{
    \footnotesize
    Probabilistic multisensory cue integration via conductance-based dendrites.
    {\bf (a)} Proposed neuronal implementation.
    Each dendritic compartment encodes a Gaussian density via the respective membrane potential $u$ (evidence) and conductance $g$ (reliability).
    While the membrane potential encodes the amount of evidence for the neurons' preferred feature, the conductance encodes the reliability of the evidence. 
    The somatic compartment represents a product model of the dendritic distributions.
    {\bf (b)} Experimental setup \citep[cf.][]{nikbakht2018supralinear}.
    Using visual and/or tactile information a rat estimates the orientation of a grating and classifies it as either vertical or horizontal.
    {\bf (c)} Network model.
    From a ground truth orientation $\theta^*$ visual and tactile stimuli are sampled with modality specific noise amplitudes and presented to two population of von-Mises feature detectors.
    All feature detectors project to two multisensory cells which are trained to respond with high/low firing rates to their preferred/anti-preferred orientations.
    {\bf (d)} Trial-averaged loss of a Bayes-optimal MAP estimate (dark gray), an unweighted estimate combining visual and tactile orientations equally (light gray), the trained model with bimodal cues (red), only visual (blue) and only tactile cues (green).
    Light colored bars indicate loss before training, light gray line denotes chance level.
    {\bf (e)} Somatic membrane potential dynamics generated by the subsequent presentation of three different orientations.
    From anti-preferred ($85^\circ$), over non-preferred ($167^\circ$) to preferred orientation ($275^\circ$).
    Fluctuations in the membrane potential reflect the reliability of the combined estimate encoded in the somatic conductance $\gtotsoma$.
  }
  \label{fig:results}
\end{figure*}

\section{Neural implementation of probabilistic cue integration}
We consider a probabilistic description of membrane potentials in multi-compartment models.
Individual dendritic compartments represent Gaussian densities via their membrane potential (mean) and total membrane conductance (precision)  (Fig.~\ref{fig:results}{\bf a}, green and blue).
The soma computes a product-of-experts model \citep{hinton2002training} of the dendritic distributions (Fig.~\ref{fig:results}{\bf a}, red).
In the present cue integration context we refer to the mean as "evidence" and the precision as "reliability".
Membrane potentials and conductances are decoupled, i.e., one can vary the membrane potential independently of the membrane conductance and vice versa, by considering parallel projections of each afferent via direct excitation and feedforward inhibition \citep{isaacson2011inhibition}.
Under the assumption of small dendritic capacitances and strong coupling of the dendritic compartments to the soma, this leads to the following somatic membrane potential distribution for a given weight matrix $\w$ and presynaptic activity $\rate$:
\begin{align}
\label{eq:soma-density}
p(u | \w, \rate) = \frac{1}{Z} e^{-\frac{\gtotsoma(\w, \rate)}{2}(u - \utotsoma(\w, \rate))^2} \; .
\end{align}
Here $\utotsoma$ is a convex combination of leak potential and dendritic membrane potentials weighted with their respective conductances and $\gtotsoma$ is a sum of leak and dendritic conductances.
From eq.~(\ref{eq:soma-density}) we obtain neuron dynamics by requiring that the somatic potential
minimizes the energy $E(u, \w, \rate) := -\log p(u|\w\rate)$ via gradient descent \citep{rao1999predictive,scellier2017equilibrium,sacramento2018dendritic}:
\begin{align}
    \cm \dot \usoma =& \gLsoma (\EL - \usoma) + \sum_{y=1}^D \gtotden_y (\utotden_y - \usoma) \; .
\end{align}
In the stationary state, the somatic potential is equal to a weighted combination of the dendritic potentials, similar to the MAP estimate in cue integration scenarios \citep{knill2004bayesian}.
The biophysics of multi-compartment neuron models with conductance-based synapses hence naturally implement an important probabilistic computation.
This makes our model particularly fitting to mixed-signal neuromorphic systems featuring conductance-based interactions \citep{schemmel2010wafer}.
A plasticity rule is obtained by requiring that the Kullback-Leibler divergence $D_\text{KL}$ between distribution of somatic potentials and some target distribution is minimized:
\begin{align}
    \Delta w_{ij}^{\text{E/I}} = \eta \left[ (\uti{i} - \utotsomai{i})(E^\text{E/I} - \utotsomai{i}) - \frac{1}{2}\left( (\uti{i} - \utotsomai{i})^2 - \frac{1}{\gtotsomai{i}} \right) \right] r_j \; ,
\end{align}
where $\ut_i$ represents a cell-specific sample from the target distribution that can be provided externally \citep{urbanczik2014learning}.
While the first part is a standard error-correcting term \citep{widrow1960adaptive,urbanczik2014learning}, the second term arises due to our probabilistic ansatz and performs reliability assignment to the respective projections.

\section{Learning a probabilistic multisensory integration task}
To illustrate our model we apply it to a probabilistic multisensory integration task: the orientation (horizontal or vertical) of a grating is to be estimated from noisy visual (V) and tactile (T) information (fig.~\ref{fig:results}{\bf b}; \citep{nikbakht2018supralinear}).
For simplicity each modality is represented by a homogeneous population of von-Mises feature detectors \citep{herz2017periodic} projecting to two multisensory cells.
From a ground truth orientation $\theta^*$, two modality specific orientations $\theta^\text{V}, \theta^\text{T}$ are sampled with modality-specific noise amplitudes ($\sigma_\text{V} << \sigma_\text{T}$) and presented to the respective feature detectors (fig.~\ref{fig:results}{\bf c}).
The two outputs are trained to signal whether the orientation is larger and smaller than $45^\circ$, respectively.
We compare the performance of the Bayes-optimal maximum-a-posteriori estimate, the naive estimate that equally weights visual and tactile stimuli, and the estimate obtained from the multisensory cells providing only visual/tactile, or both visual and tactile input (figure\ref{fig:results}{\bf c}).
While the naive and single modal estimates perform significantly worse than the MAP estimate, the multisensory estimate achieves similar error levels demonstrating that the network has successfully learned the relative reliability of the two information streams and makes use of them to integrate visual and tactile stimuli.

Despite being trained explicitly only for this task, many aspects of experimental observations are reproduced naturally by the resulting network, e.g., tight coupling and stimulus-specific tuning of excitation and inhibition \citep{isaacson2011inhibition}, stimulus-specific target potentials \citep{crochet2011synaptic,sachidhanandam2013membrane}, the stimulus-driven quenching of variability \citep{churchland2010stimulus}, sub/supra/linear multisensory neural responses \citep{perrault2005superior}, the principle of inverse effectiveness \citep{meredith1983interactions}, or reliability-dependent multisensory tuning \citep{morgan2008multisensory}.

\section{Conclusion}
Our model provides a parsimonious implementation of\linebreak Bayes-optimal cue integration in single neurons by relying on the natural dynamics of multi-compartment neurons with conductance-based dendrites.
The associated plasticity rule allows circuits to learn in a supervised setting not just to reduce output errors, but to also assign the correct relative reliabilities to different information streams.
Similar to previous models \citep{ohshiro2011normalization}, a divisive normalization operation \citep{carandini2012normalization} is a critical component.
The conductance-based nature of synaptic coupling hence may not be purely an artifact of the biological substrate, but rather enable single neurons to perform important probabilistic computations previously thought to be realized only at the circuit level \citep{ohshiro2011normalization}.
In this view, the experimental observations in multisensory integration experiments are signatures of ongoing probabilistic computations.

\begin{acks}
We gratefully acknowledge funding from the European Union, under grant agreements 604102, 720270, 785907 (HBP) and the Manfred St{\"a}rk Foundation.
\end{acks}

\bibliographystyle{ACM-Reference-Format}
\bibliography{cuecomb-nice-2020}

\appendix

\end{document}